%% file: main.tex
\documentclass[aps,reprint,nofootinbib]{revtex4-1}

\usepackage[english]{babel}
\usepackage[utf8x]{inputenc}
\usepackage[T1]{fontenc}

\usepackage{tikz}
\usepackage{amsmath}
\usepackage{hyperref}
\usepackage{cleveref}
\usepackage{graphicx}
\usepackage{booktabs}
\usepackage[acronym]{glossaries}

\usepackage{physics}
\usepackage{siunitx}

\usepackage{soul}

\input{commands}
\input{acronyms}

\begin{document}

%% Preamble
% \title{Adapting LISA time-delay interferometry to account for the Doppler effect in frequency data}

\title{Adapting time-delay interferometry for LISA data in frequency}

\author{Jean-Baptiste Bayle}
\affiliation{Jet Propulsion Laboratory, California Institute of Technology, \\ 4800 Oak Grove Drive, Pasadena, CA 91109, USA}

\author{Olaf Hartwig, Martin Staab}
\affiliation{Max-Planck-Institut für Gravitationsphysik (Albert-Einstein-Institut), \\ Callinstraße 38, 30167 Hannover, Germany}

\date{\today}

\pacs{}

\keywords{LISA; gravitational waves; orbits; noise reduction; TDI; simulation; Doppler}

%% Abstract
\begin{abstract}
\Gls{tdi} is a post-processing technique used to reduce  laser noise in heterodyne interferometric measurements with unequal armlengths, a situation characteristic of space gravitational detectors such as the \gls{lisa}. This technique consists in properly time-shifting and linearly combining the interferometric measurements in order to reduce the laser noise by several orders of magnitude and to detect gravitational waves. In this communication, we show that the Doppler shift due to the time evolution of the armlengths leads to an unacceptably large residual noise when using interferometric measurements expressed in units of frequency and standard expressions of the \gls{tdi} variables. We also present a technique to mitigate this effect by including a scaling of the interferometric measurements in addition to the usual time-shifting operation when constructing the \gls{tdi} variables. We demonstrate analytically and using numerical simulations that this technique allows one to recover standard laser noise suppression which is necessary to measure gravitational waves.
\end{abstract}
\maketitle

%% Document Content
\input{content}

%% References
\bibliographystyle{apsrev4-1}
\bibliography{references}

\end{document}

%% file: commands.tex
\renewcommand{\qc}{\,\text{,}}
\newcommand{\qs}{\,\text{.}}

\newcommand{\psd}[1]{\opbraces{S_{#1}}}

\newcommand{\delay}[1]{{\mathbf{D}_{#1}}}
\newcommand{\dotdelay}[1]{{\dot{\mathbf{D}}_{#1}}}

\DeclareSIUnit\year{yr}

%% file: acronyms.tex
\newacronym{gw}{GW}{gravitational wave}
\newacronym{esa}{ESA}{European Space Agency}
\newacronym{lisa}{LISA}{Laser Interferometer Space Antenna}
\newacronym{inrep}{INREP}{initial noise-reduction pipeline}
\newacronym{tdi}{TDI}{time-delay interferometry}

\newacronym{tt}{TT}{travel time}
\newacronym{ppr}{PPR}{proper pseudo-range}

\newacronym{psd}{PSD}{power spectral density}
\newacronym{asd}{ASD}{amplitude spectral density}

%% file: content.tex
\section{Introduction}
\label{sec:introduction}

The first observation of \glspl{gw} by the LIGO and Virgo collaborations~\cite{Abbott:2017xlt} marked the beginning of \gls{gw} astronomy. It was quickly followed by many more detections~\cite{LIGOScientific:2018mvr}. However, inherent sources of noise in ground-based detectors limit the observed frequency band to above \SI{10}{\hertz}, excluding many interesting sources, among which super-massive black hole binaries, extreme mass-ratio inspirals, or hypothetical cosmic strings. Several projects of space-borne detectors are put forward in the hope to detect \glspl{gw} in the \si{\milli\hertz} band.

One such project is the ESA-led \gls{lisa} mission~\cite{Audley:2017drz}. \Gls{lisa} aims to fly three spacecraft in a \num{2.5}-million-kilometer triangular formation, each of which exchanges laser beams with the others. The phases are monitored using sub-\si{\pico\meter} precision heterodyne interferometry, such that phase shifts induced by passing \glspl{gw} can be detected.

Laser frequency fluctuations will be the dominant source of noise, many order of magnitude above the expected level of \glspl{gw} signals~\cite{Audley:2017drz}. \Gls{tdi} is an offline technique proposed to reduce, among others, laser noise to acceptable levels~\cite{Giampieri:1996aa,Armstrong:1999hp,Tinto:1999yr,Tinto:2021aa}. It is based on the idea that the same noise affects different measurements at different times; by time-shifting and recombining these measurements, it is possible to reconstruct laser noise-free virtual interferometric signals in the case of a static constellation. We call these laser noise-free combinations the first-generation \gls{tdi} variables~\cite{Tinto:2002de,shaddock:2003aa,*Cornish:2003aa}. The algorithm has been extended to account for a breathing constellation to first order, giving rise to the so-called second-generation \gls{tdi} variables~\cite{Tinto:2004aa}. Several laboratory optical bench experiments and numerical studies have confirmed that second-generation combinations can suppress laser noise down to sufficient level to detect and exploit \glspl{gw}~\cite{Schwarze:2018lvl,Schwarze:2018lvl,Otto:2015erp,Laporte:2017bv,Cruz:2006js,Vallisneri:2005ca,Petiteau:2008ke,Bayle:2018hnm}.

In \gls{lisa}, the physical units used to represent, process, and deliver data remain to be chosen. Several studies are ongoing to determine the pros and cons of using either phase, frequency, or even chirpiness\footnote{Chirpiness is defined as the derivative of frequency.}. These include studies of the phasemeter design\footnote{Representing the variables in phase or frequency impacts most phasemeter internal processing steps, e.g., the bit depth required not to be limited by numerical quantization noise or whether or not filters must account for phase-wrapping. A detailed study of these trade-offs is beyond the scope of this paper.}, telemetry bandwidth, and potential impacts on offline noise reduction techniques, such as \gls{tdi}.

Most \gls{tdi} studies indifferently assume that the measurements are expressed in terms of interferometric beatnote phases or frequencies \cite{Tinto:2021aa}. However, these studies disregard the Doppler shifts that arise when using units of frequency~\cite{Tinto:2021aa,Vallisneri:2005ca,Petiteau:2008ke,Bayle:2018hnm}: the relative motion of the spacecraft induces time-varying frequency shifts in the beatnote frequencies, which reduce the performance of standard \gls{tdi} algorithms. In fact, as we show below, the standard formulation of \gls{tdi} applied to frequency data no longer suppresses laser noise to the required level. We however demonstrate that these \gls{tdi} algorithms can be easily modified to account for Doppler shifts when using frequency data. Ultimately, we recover the same laser noise-reduction performance as one obtains when using units of phase.
% These studies evaluate, for example, the implications of using 

% Technical arguments: 
%  - phase is very strong ramp, regulary overflowing -> this impacts filter design in the phasemeter. Time interval between overflows is ultimately limited by the variable size (telemetry budget). 
%  - phase in principle contains more information

% The heterodyne beatnotes in \gls{lisa} typically have frequencies of \SI{5}{\mega\hertz} to \SI{25}{\mega\hertz}, such that the corresponding total phase is a rapidly evolving linear function. It will therefore overflow at regular time intervals when represented by a fixed precision binary variable, where the timespan between overflows is determined by. This can have technical implications for the hardware design, since any onboard filters need to account for the corresponding large phase jumps. 

% In addition to impacting the onboard processing and transmission of data, the choice between phase and frequency also has an impact on offline processing algorithms such as \gls{tdi}.

The paper is structured as follows: in \cref{sec:interferometric-measurements}, we derive the expression of the interferometric measurements in terms of frequency and show how Doppler shifts couple. Then, in \cref{sec:residual-in-tdi}, we evaluate the additional noise due to these Doppler shifts in the \gls{tdi} variables and show that it does not meet the requirements. A procedure to mitigate this effect is presented in \cref{sec:doppler-tdi}. We show that the Doppler couplings can be reduced to levels below the requirements, and confirm the analytical study by numerical simulations in \cref{sec:simulation}. Finally, we conclude in \cref{sec:conclusion}.

\section{Interferometric measurements}
\label{sec:interferometric-measurements}

In this paper, we follow the latest recommendations on conventions and notations established by the \gls{lisa} Consortium. Since these conventions are relatively new, we provide in \cref{sec:convention-mapping} a mapping between the various existing conventions.

\begin{figure}
    \centering
    \includegraphics[width=\columnwidth]{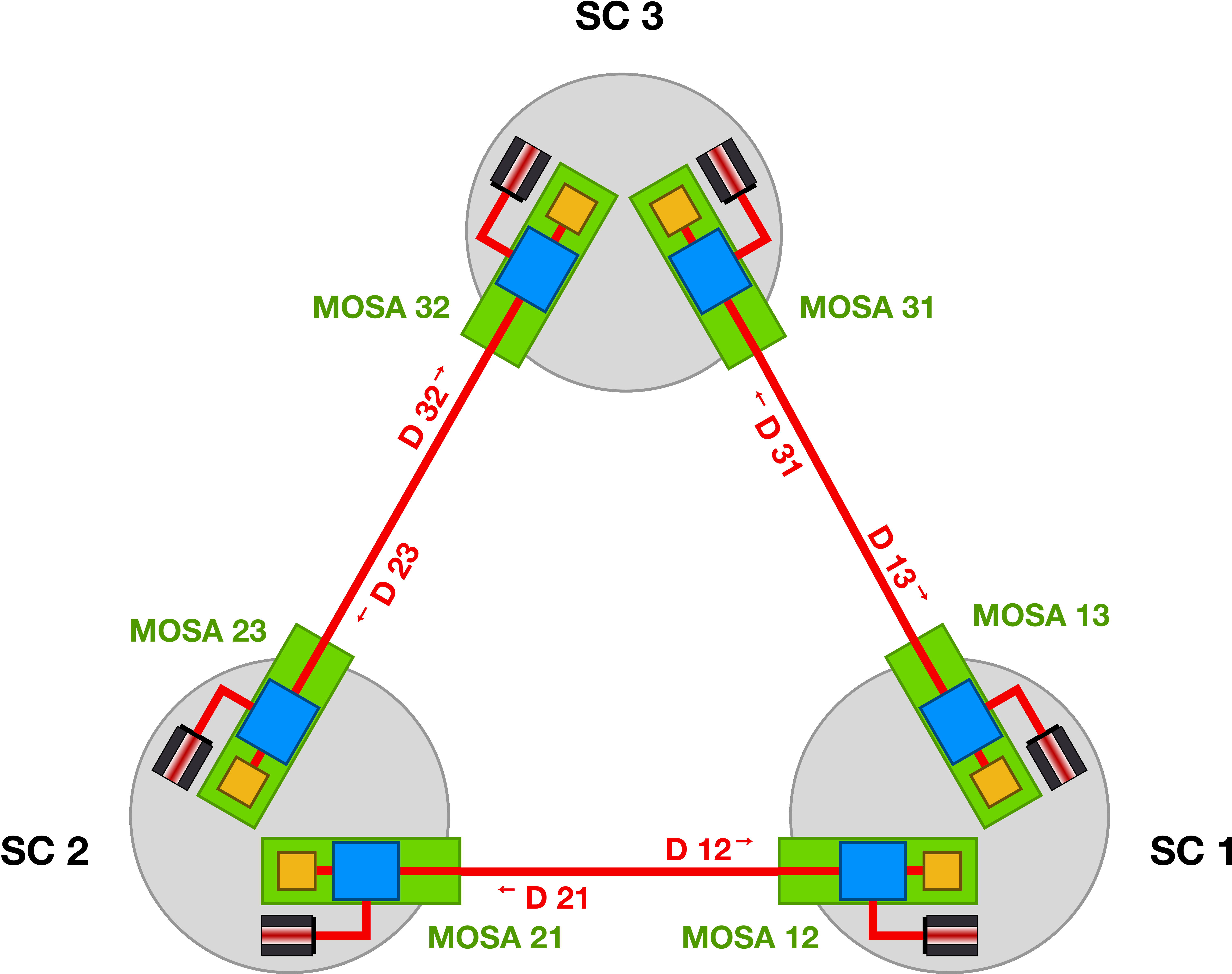}
    \caption{Labelling conventions used for spacecraft, light \glspl{tt}, lasers, optical benches, and interferometric measurements.}
    \label{fig:labelling}
\end{figure}

We label the spacecraft as presented in \cref{fig:labelling}. The optical benches are labelled with two indices $ij$. The former matches the index $i$ of the spacecraft hosting the optical bench, while the second index is that of the spacecraft $j$ exchanging light with the optical bench. Any subsystem or measurement uniquely attached to an optical bench share the same indices.

As an example, the light travel time (\gls{tt}) measured on optical bench~$ij$ represents the time of flight of a photon received by spacecraft $i$ and emitted from spacecraft $j$. Note the unusual ordering of the indices (\textit{receiver}, \textit{emitter}); while this choice may seem peculiar at first, it will turn out most useful when writing \gls{tdi} equations in \cref{sec:residual-in-tdi} and later.

We assume that the spacecraft follow perfectly the test masses they host; therefore, their orbits are described as geodesics around the Sun. Accounting for the sole influence of the Sun, the computation of their positions and velocities reduces to a two-body problem, which can be solved semi-analytically~\cite{Dhurandhar:2004rv,Nayak:2006zm}. A more realistic approach uses a set of orbits computed using numerical integration (which includes the influence of the more massive objects in the Solar System) optimized for a given set of constraints, such as minimizing the motion of the spacecraft relative to one another\cite{Nayak:2006zm,Joffre:2020,Martens:2021phh}.

From these orbits, one can compute the light \gls{tt}, denoted by $d_{ij}$, of a photon received by spacecraft $i$ and emitted from spacecraft $j$. Because no sets of orbits ensures a static constellation, we say that the constellation breathes. A direct consequence of this is that the light \glspl{tt} changes with time, and we write, e.g., $d_{ij}(t)$.

Each spacecraft contains, among others, two laser sources and two optical benches, labelled according to \cref{fig:labelling}. Three interferometric signals, namely the inter-spacecraft\footnote{Formerly known as the science or long-arm interferometer.} $\text{isc}_{ij}(t)$, test-mass $\text{tm}_{ij}(t)$, and reference $\text{ref}_{ij}(t)$ beatnotes, are measured on each optical bench $ij$~\cite{Audley:2017drz}. In addition, a pseudo-random code is used to modulate the laser beams exchanged by the spacecraft \cite{Heinzel:2011aa,jose-esteban:2011aa}. The signal is then correlated with a local version to provide an estimate of the light \glspl{tt}, called measured pseudo-ranges. Various errors entering the measured pseudo-ranges and their impact on data processing and analysis are the focus of ongoing studies \cite{Wang:2014aa,*Wang:2015aa}. We shall assume here that the measured pseudo-ranges furnish perfect measurements of the light \glspl{tt}, and therefore, we shall use indifferently pseudo-ranges or light \glspl{tt}, both denoted $d_{ij}(t)$.

Moreover, we will assume here that each spacecraft contains only one laser, which is used in both optical benches. This is without loss of generality, since this situation can be achieved in practice either by locking the two lasers on board each spacecraft\footnote{The precise locking configuration is still under study.} or by constructing the intermediary variables $\eta$~\cite{Tinto:2002de,Tinto:2021aa}.

On board spacecraft~$i$, the phase of the local laser beam in units of cycles is denoted $\Phi_i(t)$. It contains the phase ramp due to the average laser frequency (around $\SI{281}{\tera\hertz}$), as well as small in-band phase fluctuations, dominated by the instability of the reference cavity used for stabilization (around $\SI{30}{\hertz\per\sqrt\hertz}$ when expressed as a frequency noise~\cite{Audley:2017drz}).

The phase of the beam emitted by spacecraft $j$ and received on $i$ at time $t$ reads
\begin{equation}
    \Phi_{i \leftarrow j}(t) = \Phi_j(t - d_{ij}(t) - H_{ij}(t))
    \qc
    \label{eq:distant-beam-phase-explicit-full}
\end{equation}
where $d_{ij}(t)$ is the light \gls{tt} between $j$ and $i$ without any \glspl{gw}. The effect of passing \glspl{gw} are modelled by an additional delay $H_{ij}(t)$. Because this quantity is very small with respect to $d_{ij}(t)$, we Taylor-expand the phase to write $H_{ij}(t)$ as an independent term, and get
\begin{equation}
    \Phi_{i \leftarrow j}(t) = \Phi_j(t - d_{ij}(t)) - \nu_j(t - d_{ij}(t)) H_{ij}(t)
    \qs
    \label{eq:distant-beam-phase-explicit}
\end{equation}

For more the sake of clarity, we drop the time dependence and introduce the delay operator $\delay{ij}$, defined by
\begin{equation}
    \delay{ij} x(t) = x(t - d_{ij}(t))
    \qc
\end{equation}
for any signal $x(t)$. We shall also use the compact notation for chained delay operators, formally defined by 
\begin{equation}
    \delay{i_1 i_2 \dots i_n} = \delay{i_1 i_2} \delay{i_2 i_3} \dots \delay{i_{n-1} i_n}
    \qc
\end{equation}
such that we have, \textit{e.g.}, in the case of two delay operators,
\begin{equation}
\begin{split}
    \delay{ijk} x(t) &= \delay{ij} \delay{jk} x(t) = \delay{ij} x(t - d_{jk}(t)) \\ 
    &= x\Big(t - d_{ij}(t) - d_{jk}\big(t - d_{ij}(t)\big)\Big)
    \qs
\end{split}
\end{equation}

Using these conventions, \cref{eq:distant-beam-phase-explicit} becomes
\begin{equation}
    \Phi_{i \leftarrow j} = \delay{ij} \Phi_j - (\delay{ij} \nu_j) H_{ij}
    \qs
    \label{eq:distant-beam-phase}
\end{equation}

The frequency of the local laser beam on optical bench~$ij$ is simply the derivative of the total phase $\nu_i = \dot \Phi_i$. Similarly, the frequency of the distant beam is obtained by Taylor-expanding the derivative of \cref{eq:distant-beam-phase-explicit-full},
\begin{equation}
\begin{split}
    &\nu_{i \leftarrow j}(t) = \dot \Phi_{i \leftarrow j}(t) = [1 - \dot{d}_{ij}(t) - \dot H_{ij}(t)] \\
    &\qquad \times [\nu_j(t - d_{ij}(t)) - \dot{\nu}_j(t - d_{ij}(t)) H_{ij}(t)]
    \qs
\end{split}
\end{equation}

\begin{figure*}
    \centering
    \includegraphics[width=\textwidth]{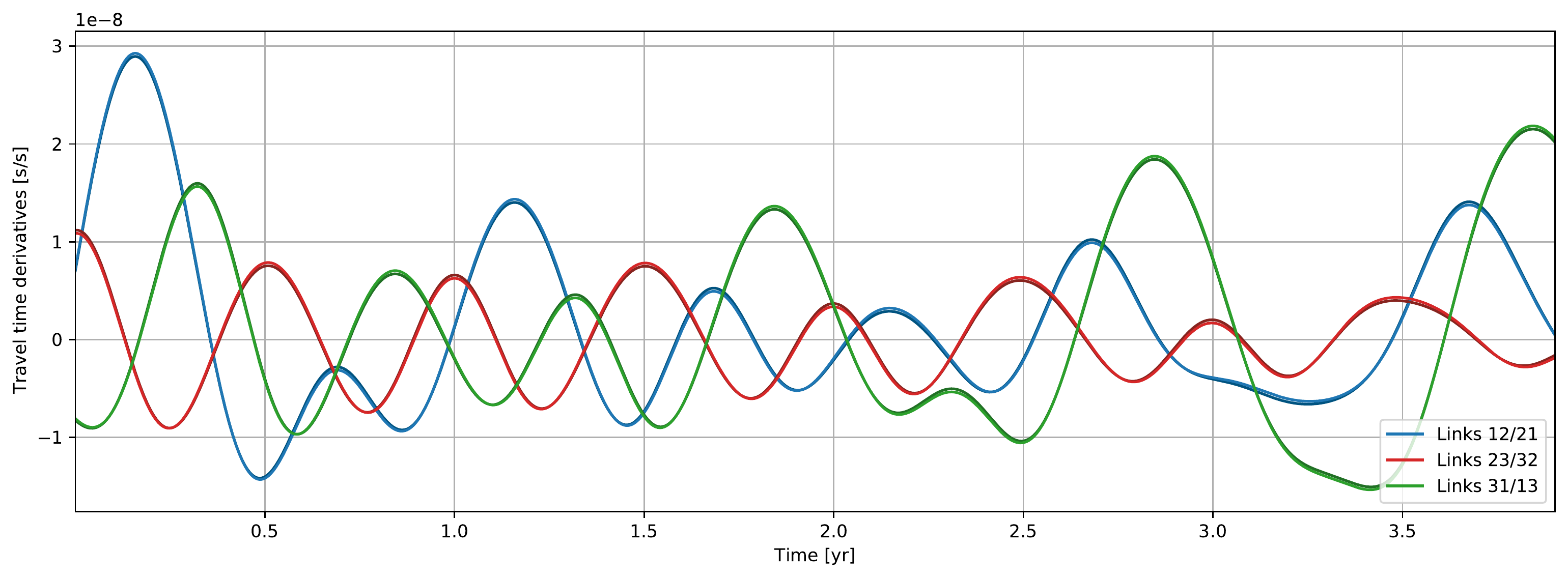}
    \caption{Light travel time derivatives for realistic orbits.}
    \label{fig:orbits}
\end{figure*}

In the following, we neglect all terms in $\dot{\nu}_j H_{ij}$. Indeed, the rate of change of $\nu_j$ is driven by laser noise\footnote{We expect that laser frequencies also vary due to the frequency plan, by \si{\mega\hertz} over the timescale of months. This yields terms of the same order of magnitude, so that our reasoning holds.}. Using the expected level of laser noise and integrating it over the \gls{lisa} frequency band, $\dot{\nu}_j \approx \SI{E2}{\hertz\per\second}$. Therefore, $\dot{\nu}_j H_{ij} \approx \SI{E-18}{\hertz} \ll \nu_j \dot{H}_{ij} \approx \SI{E-7}{\hertz}$. Dropping the time dependence and using our delay operator,
\begin{equation}
    \nu_{i \leftarrow j} = (1 - \dot d_{ij}) \delay{ij} \nu_j - (\delay{ij} \nu_j) \dot H_{ij}
    \qs
    \label{eq:distant-beam-freq}
\end{equation}
The factor $\dot d_{ij}(t) \delay{ij} \nu_j$ is often referred to as the \textit{Doppler shift}, and is proportional to the time derivative of the light \gls{tt}. \Cref{fig:orbits} show the time variations of such quantities for realistic orbits \cite{Joffre:2020,Martens:2021phh}, of the order of \num{E-8} (or \SI{3}{\meter\per\second}).

The inter-spacecraft interferometer mixes the local and distant beams. The beatnote phase $\Phi^\text{isc}_{ij}$ can easily be expressed as the difference of the beam phases,
\begin{equation}
    \Phi^\text{isc}_{ij} = \Phi_{i \leftarrow j} - \Phi_i
    = \delay{ij} \Phi_j - \Phi_i - (\delay{ij} \nu_j) H_{ij}
    \qs
    \label{eq:isc-phase}
\end{equation}
In units of frequency, we have
\begin{equation}
\begin{split}
    &\nu^\text{isc}_{ij} = (1 - \dot d_{ij}) \delay{ij} \nu_j - \nu_i - (\delay{ij} \nu_j) \dot H_{ij}
    \qc
    \label{eq:isc-freq}
\end{split}
\end{equation}
where the term $\dot d_{ij} \delay{ij} \nu_j$ is the Doppler shift.

In \cref{eq:isc-freq}, the main in-band contribution is laser noise, which does not cancel out\footnote{Even if lasers are locked such that there is only one laser noise, it is not sufficiently suppressed due to the large delays.} and remains orders of magnitude above the gravitational-wave signal $(\delay{ij} \nu_j) \dot H_{ij} \approx \SI{E-7}{\hertz}$. In order to detect and extract gravitational information from the measurements, laser noise must be reduced by at least 8 orders of magnitude.

% nu0 = 282THz = 2.8e14
% GW 1e-21
% GW * nu0 = 1e-21*2.8e14 = 2.8e-7
% laser noise 30 Hz -> E1 -> 8 orders of magnitudes
% doppler 1e-8  -> noise * doppler = 3e-7

\section{Residual noise due to Doppler shifts in TDI}
\label{sec:residual-in-tdi}

\Gls{tdi} is a technique proposed to reduce instrumental noises, including laser noise, to acceptable levels. The starting point for the main \gls{tdi} algorithm is usually to compute the so-called intermediary variables $\xi$ and $\eta$, which are used to remove spacecraft jitter noise and reduce the number of lasers to three. While we already consider only one laser per spacecraft, we will further neglect spacecraft jitter noise, such that we can directly write $\eta_{ij} = \Phi^\text{isc}_{ij}$ in phase, or $\eta_{ij} = \nu^\text{isc}_{ij}$ in frequency.

The next step is to reduce laser noise. Several laser noise-reducing combinations have been proposed. E.g., the second-generation Michelson variable $X_2$ reads \cite{Tinto:2021aa}
% \begin{equation}
% \begin{split}
%     &X_1 = (1 - \delay{12} \delay{21}) \eta_{13} + (1 - \delay{12} \delay{21}) \delay{13} \eta_{31} \\
%     &\qquad - (1 - \delay{13} \delay{31}) \eta_{12} - (1 - \delay{13} \delay{31}) \delay{12} \eta_{21}
%     \qc
%     \label{eq:X2}
% \end{split}
% \end{equation}
% \begin{equation}
% \begin{split}
%     &X_2 = (1 - \delay{12} \delay{21} - \delay{12} \delay{21} \delay{13} \delay{31} \\
%     &\qquad + \delay{13} \delay{31} \delay{12} \delay{21} \delay{12} \delay{21}) (\eta_{13} + \delay{13} \eta_{31}) \\
%     &\quad - (1 - \delay{13} \delay{31} - \delay{13} \delay{31} \delay{12} \delay{21} \\
%     &\qquad + \delay{12} \delay{21} \delay{13} \delay{31} \delay{13} \delay{31}) (\eta_{12} + \delay{12} \eta_{21})
%     \qc
%     \label{eq:X2}
% \end{split}
% \end{equation}
\begin{equation}
\begin{split}
    &X_2 = (1 - \delay{121} - \delay{12131} + \delay{1312121}) (\eta_{13} + \delay{13} \eta_{31}) \\
    &\quad - (1 - \delay{131} - \delay{13121} + \delay{1213131}) (\eta_{12} + \delay{12} \eta_{21})
    \qs
    \label{eq:X2}
\end{split}
\end{equation}
The two other Michelson variables $Y_2, Z_2$ are obtained by circular permutation of the indices $1 \rightarrow 2 \rightarrow 3 \rightarrow 1$.

In the following, we shall ignore any technical reasons for imperfect laser noise reduction, such as flexing-filtering coupling~\cite{Bayle:2018hnm}, interpolation errors or ranging errors, and only consider the maximum theoretical laser noise reduction achievable.

In case of phase, we know that the residual laser noise in this variable is given by the non-commutation of delay operators~\cite{Bayle:2018hnm},
\begin{equation}\label{eq:commut}
    X^\Phi_2 = \comm{\comm{\delay{131}}{\delay{121}}}{\delay{12131}} \Phi_1
    \qs
\end{equation}
Expanding this expression to second order in the average \gls{tt} derivatives $\dot{d}$'s and first order in average \gls{tt} second derivatives $\ddot{d}$'s, and assuming that these quantities are symmetric in $i$, $j$, the difference of the delays applied to the phase $\Phi_1$ in the two terms from \cref{eq:commut} reads
\begin{equation}
    \Delta d = 8 \bar d \qty(\bar{\dot d}_{12}^2 - \bar{\dot{d}}_{31}^2) - 16 \bar{d}^2 \qty(\bar{\ddot{d}}_{12} - \bar{\ddot{d}}_{31})
    \qc
\end{equation}
where the first term matches the results of~\cite{Bayle:2018hnm}. In terms of \gls{psd}, we have
\begin{equation}
    \psd{X^\Phi_2}(\omega) = \omega^2 \Delta d^2 \psd{\Phi}(\omega)
    \qc
    \label{eq:X2-laser-psd-phase}
\end{equation}
where $\psd{\Phi}(\omega)$ is dominated by the \gls{psd} of the laser noise expressed in cycles.

Now, let us assess the impact of Doppler shifts if one uses naively the traditional second generation \gls{tdi} algorithm using measurements in units of frequency. For this, we can insert \cref{eq:isc-freq} in \cref{eq:X2}. The only structural difference between \cref{eq:isc-freq} and \cref{eq:isc-phase} is the additional Doppler term $\dot d_{ij} \delay{ij} \nu_j$. Because \gls{tdi} is a linear operation, we can immediately give the residual laser noise in terms of frequency when applying the same algorithm,
\begin{equation}
    X^\nu_2 = \comm{\comm{\delay{131}}{\delay{121}}}{\delay{12131}} \nu_1 + \delta X^\nu_2
    \qc
    \label{eq:X2-freq}
\end{equation}
where $\delta X^\nu_2$ is a function of the Doppler shifts,
\begin{equation}
\begin{split}
    \delta X^\nu_2 &= (1 - \delay{131} - \delay{13121} + \delay{1213131}) \\
    &\qquad\qquad\qquad \times (\dot d_{12} \delay{12} \nu_2 + \dot d_{21} \delay{121} \nu_1) \\
    &\quad - (1 - \delay{121} - \delay{12131} + \delay{1312121}) \\
    &\qquad\qquad\qquad \times (\dot d_{13} \delay{13} \nu_3 + \dot d_{31} \delay{131} \nu_1)
    \qs
    \label{eq:delta-X2}
\end{split}
\end{equation}

\begin{figure}
    \centering
    \includegraphics[width=\columnwidth]{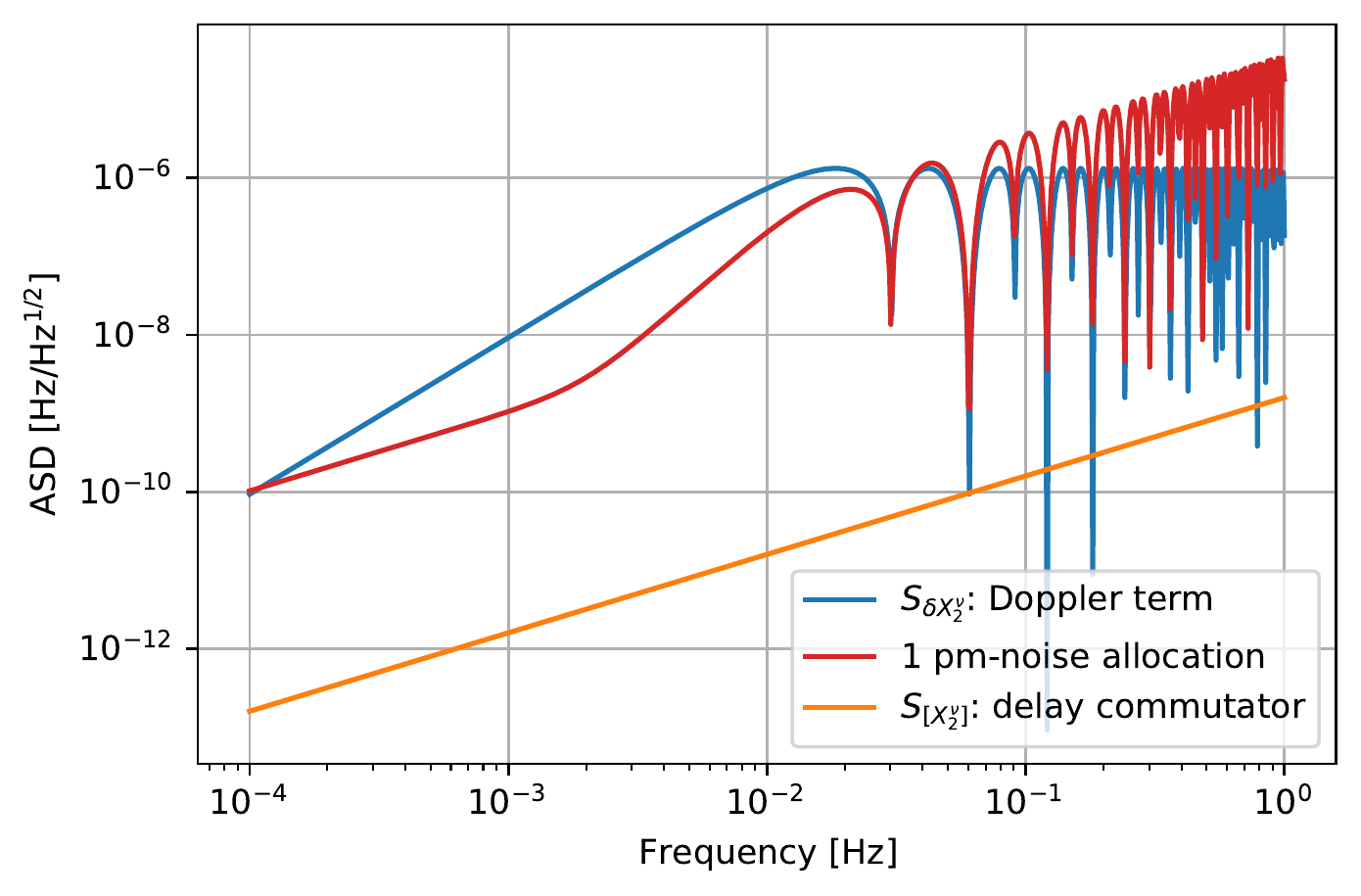}
    \caption{Amplitude spectral density of the second generation \gls{tdi} combination when using measurements expressed in units of frequency. The blue curve shows the amplitude of  Doppler-related terms, c.f.~\cref{eq:delta-X2-psd}, the orange curve shows the amplitude of the delay commutators, c.f.~\cref{eq:X2-without-dopplers}, while the red curve presents the usual \gls{lisa} \SI{1}{\pico\meter}-noise allocation, c.f.~\cref{eq:noise-allocation}. The light travel times used in this simulation are presented in \cref{fig:orbits}.}
    \label{fig:analytical-curves}
\end{figure}

A rough estimation of this Doppler coupling can be computed from $\delta X^\nu_2 \approx \bar{\dot d} \nu$. Plugging orders of magnitudes for the \glspl{tt} derivatives and laser noise yields a Doppler coupling at $\SI{E-6}{\hertz}$, above the expected level for our \gls{gw} signals ($\SI{E-7}{\hertz}$). It is also above the level of the traditional residuals of \gls{tdi}, given by the first term of \cref{eq:X2-freq} and shown in \cref{fig:analytical-curves}. As a consequence, the \gls{psd} of the residual noise for the $X_2^\nu$ \gls{tdi} variable is dominated by the Doppler coupling,
\begin{equation}\label{eq:PSD-X2-freq}
   \psd{ X^\nu_2}(\omega) \approx \psd{\delta X^\nu_2}(\omega)
   \qs
\end{equation}

Assuming that all laser frequencies are uncorrelated, a more precise computation yields the \gls{psd} of this extra residual noise,
\begin{equation}
\begin{split}
    \psd{\delta X^\nu_2}(\omega) &\approx 16 \psd{\nu} \sin[2](\omega \bar d) \sin[2](2 \omega \bar d)
    \\
    &\qquad \times \qty(\bar{\dot d}_{12}^2 + \bar{\dot d}_{31}^2 + (\bar{\dot d}_{12} - \bar{\dot d}_{31})^2) 
    \qs
    \label{eq:delta-X2-psd}
\end{split}
\end{equation}

This is to be compared with the residual laser noise in terms of frequency when one disregards Doppler effects. It is given by replacing $\psd{\Phi}$ with $\psd{\nu}$ in \cref{eq:X2-laser-psd-phase},
\begin{equation}
    \psd{[X^{\nu}_2]}(\omega)= \omega^2 \Delta d^2 \psd{\nu}(\omega)
    \qs
    \label{eq:X2-without-dopplers}
\end{equation}

In \cref{fig:analytical-curves}, we show those analytical curves alongside the usual \gls{lisa} Performance Model's \SI{1}{\pico\meter}-noise allocation curve, given by
\begin{equation}
\begin{split}
    &\psd{X_2^\text{alloc}}(\omega) = 64 \omega^2 \sin[2](\omega \bar d) \sin[2](2 \omega \bar d)  \\
    &\qquad \times \qty(\frac{\SI{1}{\pico\meter\hertz^{-1/2}}}{\lambda})^2 \qty[1 + \qty(\frac{\SI{2e-3}{\hertz}}{\omega / 2 \pi})^4]
    \qs
    \label{eq:noise-allocation}
\end{split}
\end{equation}
The extra residual laser noise due to Doppler terms is above or at the same level as the \gls{gw} signal, and far above the usual laser noise residual when one disregards the Doppler effect. Therefore, a procedure to mitigate this effect is required if one wishes to use frequency measurements.

\section{Adapting time-delay interferometry for Doppler shifts}
\label{sec:doppler-tdi}

As mentioned in the previous section, accounting for the Doppler effect in the inter-spacecraft beatnote frequency comes down to replacing the delay operator $\delay{ij}$ in \cref{eq:isc-phase} by $(1 - \dot d_{ij}) \delay{ij}$. We can formalize it by introducing the Doppler-delay operator,
\begin{equation}
    \dotdelay{ij} = (1 - \dot d_{ij}) \delay{ij}
    \qc
    \label{eq:doppler-delay}
\end{equation}
such that laser noise entering \cref{eq:isc-freq} takes the same algebraic form as its phase counterpart \cref{eq:isc-phase},
\begin{equation}
    \nu^\text{isc}_{ij} = \dotdelay{ij} \nu_j - \nu_i + (\delay{ij} \nu_j) \dot H_{ij}
    \qs
    \label{eq:isc-freq-doppler-delay}
\end{equation}

We now introduce a new type of second generation \gls{tdi} combination by considering the standard expression from \cref{eq:X2} but using the Doppler-delay operators introduced in \cref{eq:doppler-delay}. The new \gls{tdi} variable writes 
\begin{equation}
\begin{split}
    &\dot{X}_2 = (1 - \dotdelay{121} - \dotdelay{12131} + \dotdelay{1312121}) (\eta_{13} + \dotdelay{13} \eta_{31}) \\
    &\quad - (1 - \dotdelay{131} - \dotdelay{13121} + \dotdelay{1213131}) (\eta_{12} + \dotdelay{12} \eta_{21})
    \qs
    \label{eq:X2-for-dopplers}
\end{split}
\end{equation}
The algebraic form of this expression is now identical in phase and frequency, and we immediately recover the residual noise given in \cref{eq:X2-laser-psd-phase},
\begin{equation}
    \dot{X}^\nu_2 = \comm{\comm{\dotdelay{131}}{\dotdelay{121}}}{\dotdelay{12131}} \nu_1
    \qs
\end{equation}
A direct comparison with \cref{eq:X2-freq} demonstrates that the new \gls{tdi} variable introduced in \cref{eq:X2-for-dopplers} is not impacted by the Doppler noise $\delta X_2^\nu$.

To compute the \gls{psd} of the $\dot X_2^\nu$ residual laser noise, we study the commutator of Doppler-delay operators
\begin{equation}
    y = \comm{\dotdelay{i_1 j_1} \dots \dotdelay{i_n j_n}}{\dotdelay{k_1 l_1} \dots \dotdelay{k_n l_n}}
    \qs
    \label{eq:doppler-delay-commutator}
\end{equation}
As one can observe in \cref{fig:orbits}, the light \gls{tt} derivatives evolve slowly with time, with $\ddot d \Delta t \sim 10^{-14} \ll \dot d \sim 10^{-8}$ if $\Delta t \sim \SI{10}{\second}$ is the timescale of the \glspl{tt} considered here. Therefore, we can assume that $\dot d$'s are constant when computing $y$. \Cref{eq:doppler-delay-commutator} can then be factored as
\begin{equation}
\begin{split}
    y = \qty(\prod_{m=1}^n{(1-\dot{d}_{i_m j_m})})\qty(\prod_{m=1}^n{(1-\dot{d}_{k_m l_m})}) \times \\
    \comm{\delay{i_1 j_1} \dots \delay{i_n j_n}}{\delay{k_1 l_1} \dots \delay{k_n l_n}}
    \qs
\end{split}
\end{equation}

The factor that contains the \gls{tt} derivatives is a constant, which, to first order, deviates from 1 by $2 \bar{\dot d}n \approx \num{E-7}$. We can therefore neglect it when estimating the \gls{psd}. For this reason, the \gls{psd} of the laser noise residual for the new \gls{tdi} variable introduced in \cref{eq:X2-for-dopplers} is then given by
\begin{equation}\label{eq:psd-dot-X2}
    \psd{ \dot X^\nu_2}(\omega)=\psd{[X^{\nu}_2]}(\omega)\, ,
\end{equation}
whose expression is explicitly given in \cref{eq:X2-without-dopplers}. A direct comparison with \cref{eq:PSD-X2-freq} shows that the \gls{psd} of the new ${\dot X}_2^\nu$  \gls{tdi} variable is not impacted by the unacceptably large contribution from $\delta X_2^\nu$.

The method presented in this section which consists in replacing $\delay{ij}$ by $\dotdelay{ij}$ in the usual \gls{tdi} combinations in order to remove the effect of Doppler shift is very general and can be applied to any \gls{tdi} combination. 

\section{Simulation results}
\label{sec:simulation}

Using LISANode~\cite{BayleThesis} and \texttt{lisainstrument}, a Python simulator based on LISANode, we simulated the interferometric measurements as frequency deviations from the average beatnote frequencies. These frequency deviations include only laser noise, which is Doppler-shifted during propagation. We assumed 3 free-running lasers for this study, and used a high sampling rate, such that effects of onboard filtering appear off band. We used the same realistic orbits and light travel times as presented in \cref{fig:orbits}, and simulated \num{E7} samples, i.e., a bit less than \num{12} days.

\begin{figure*}
    \centering
    \includegraphics[width=\textwidth]{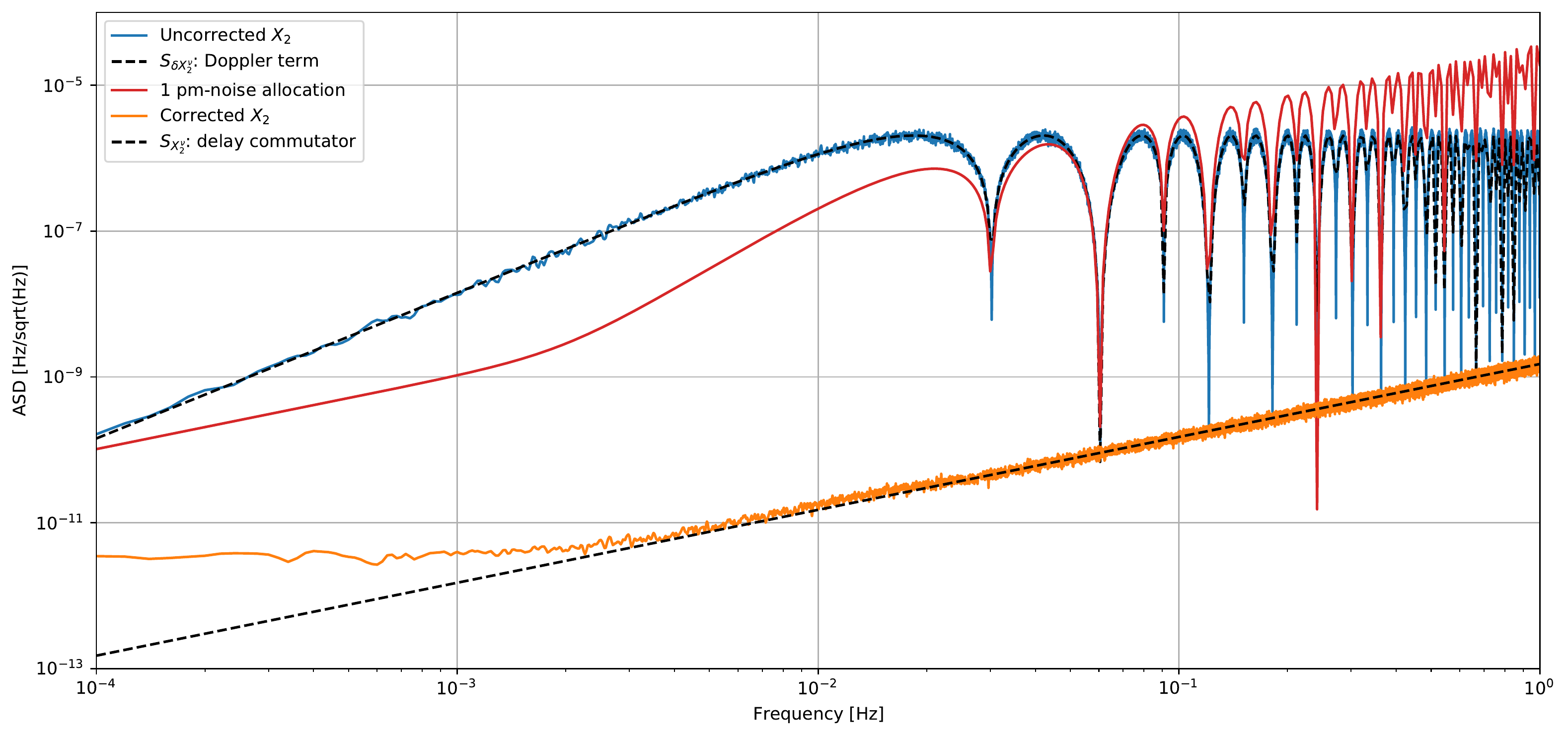}
    \caption{ Amplitude spectral density of the residual laser noise in $X_2^\nu$ obtained using data in units of frequency, with the traditional algorithms (in blue) and Doppler correction (in orange). The theoretical models from Eqs.~(\ref{eq:delta-X2-psd}) and (\ref{eq:X2-without-dopplers}) are superimposed as black dashed lines. These curves need to be compared with the $\SI{1}{\pico\meter}$-noise allocation (in red).}
    \label{fig:simulation}
\end{figure*}

The \gls{tdi} processing was performed using PyTDI. In \cref{fig:simulation}, we compare 2 different scenarios using the same input data. The blue curve shows the \gls{asd} of the residual laser noise when the standard second-generation Michelson $X^\nu_2$ variable is used. We superimpose the model for the expected excess of noise $\delta X^\nu_2$ due to Doppler effect given in \cref{eq:delta-X2-psd}, and check that it matches our simulated results. Alternatively, the orange curve shows the \gls{asd} of the residual laser noise when the Doppler-corrected second-generation Michelson $\dot{X}^\nu_2$ variable is used. It is superimposed with the analytical expectation given in \cref{eq:psd-dot-X2} in most of the band, until we reach a noise floor around \SI{2E-12}{\hertz\per\sqrt{\hertz}}. This noise floor is in agreement with the numerical accuracy typically achieved in our simulations.

These simulations confirm the analytical results developed in the previous section. In particular, it shows that the residual noise of the new \gls{tdi} variable introduced in \cref{eq:X2-for-dopplers} is similar to the one obtained with the standard \gls{tdi} combinations when the Doppler effect is neglected. Say in other words, the \gls{tdi} variable corrects efficiently for the Doppler contribution which otherwise induces an unacceptably large noise.

\section{Conclusion}
\label{sec:conclusion}

In this paper, we show that the \gls{tdi} combinations found in the literature \cite{Tinto:2021aa} do not reduce laser noise to required levels when applied to data in units of frequency and we provide an analytical formulation of the additional residual noise. We then propose a technique to adapt existing \gls{tdi} combinations to data in units of frequency. We show through analytical studies, as well as with numerical simulations that we recover the original laser-noise reduction performance, compatible with requirements to detect and exploit \glspl{gw} signals.

\Gls{tdi} is required to suppress primary noises in the interferometric measurements to levels below that of \gls{gw} signals. Existing formulations are based on the assumption that these measurements are expressed in terms of phase, or disregard the impact of Doppler shifts when data in frequency are used \cite{Tinto:2021aa}. However, applying these \gls{tdi} algorithms to data in units of frequency yields extra noise residuals due to the Doppler shift induced by the time variation of the arm lengths. This extra noise residuals are larger than \gls{gw} signals.  To account for Doppler shifts we reformulate the \gls{tdi} combinations by replacing delay operators by their Doppler equivalent, which not only shift measurement in time but also scale them by the corresponding Doppler factor, see \cref{eq:doppler-delay}. We show that this general procedure yields new \gls{tdi} combinations, whose performance when applied to measurements in frequency match that of the traditional combinations when working in units of phase.

This is a major result to study the impact of different physical units in \gls{lisa} data processing. We show that laser noise reduction can reach similar levels using phase or frequency measurements. Nevertheless, computing the \gls{tdi} using frequency measurements require the knowledge of both the \gls{tt} and their time derivatives while only the \gls{tt} are needed in order to construct \gls{tdi} variables using phase measurements. This might impact the development of a Kalman filter whose goal is to provide an estimate of the \gls{tt} \cite{Wang:2014aa,Wang:2015aa}. Finally, it is known that the clocks from the various spacecraft will drift with respect to each other because of relativistic effects \cite{pireaux:2007sh} and because of clock noise. Therefore, the \gls{lisa} pre-processing will also include a synchronization of the clocks from the 3 spacecraft \cite{Tinto:2021aa}. How this synchronization will impact the construction of \gls{tdi} variables is currently under exploration and might differ if one uses phase or frequency units. A detailed study of the interplay of \gls{tdi} with clock synchronization is left for a dedicated study. Finally, let us mention that using frequency units to perform the data analysis of \gls{lisa} may also impact the sources parameters inference since the \gls{tdi} response function used in Bayesian algorithm may have to include the currently neglected Doppler correction.

\section*{Acknowledgments}

The authors are grateful to the SYRTE Theory and Metrology Group, in particular Aurélien Hees, Marc Lilley, Peter Wolf, and Christophe Le Poncin-Lafitte, for the useful discussions and suggestions to improve the presentation of the article.

JBB was supported by an appointment to the NASA Postdoctoral Program at the Jet Propulsion Laboratory, California Institute of Technology, administered by Universities Space Research Association under contract with NASA. Part of this research was carried out at the Jet Propulsion Laboratory, California Institute of Technology, under a contract with the National Aeronautics and Space Administration (80NM0018D0004).

OH and MS gratefully acknowledge support by the Deutsches Zentrum für Luft- und Raumfahrt (DLR) with funding from the Bundesministerium für Wirtschaft und Technologie (Project Ref. Number 50 OQ 1801, based on work done under Project Ref. Number 50 OQ 1301 and 50 OQ 0601). This work was also supported by the Max-Planck-Society within the LEGACY (“Low-Frequency Gravitational Wave Astronomy in Space”) collaboration (M.IF.A.QOP18098).

\appendix
\section{Mapping between conventions}
\label{sec:convention-mapping}

The following table gives the mapping between the double-index conventions used in the article, and historic ones using primed indices, used in, e.g.,~\cite{Bayle:2018hnm,BayleThesis,Otto:2015erp,Petiteau:2008ke}. We give the correspondence for optical bench and associated subsystems and quantities, and that for light \glspl{tt} and their derivatives.

\begin{table}[h]
    \begin{tabular}{ccc}
        \toprule
        Double-index & \begin{tabular}{c}Primed indices \\ for optical benches\end{tabular} & \begin{tabular}{c}Primed indices \\ for light \glspl{tt}\end{tabular} \\
        \midrule
        12 (e.g., $\nu^\text{isc}_{12}$ or $d_{12}$) & 1 (e.g., $\nu^\text{isc}_1$) & 3 (e.g., $d_3$) \\
        23 (e.g., $\nu^\text{isc}_{23}$ or $d_{23}$) & 2 (e.g., $\nu^\text{isc}_2$) & 1 (e.g., $d_1$) \\
        31 (e.g., $\nu^\text{isc}_{31}$ or $d_{31}$) & 3 (e.g., $\nu^\text{isc}_3$) & 2 (e.g., $d_2$) \\
        13 (e.g., $\nu^\text{isc}_{13}$ or $d_{13}$) & $1^\prime$ (e.g., $\nu^\text{isc}_{1'}$) & $2^\prime$ (e.g., $d_{2'}$) \\
        32 (e.g.,  $\nu^\text{isc}_{32}$ or $d_{32}$) & $3^\prime$ (e.g., $\nu^\text{isc}_{3'}$) & $1^\prime$ (e.g.,\, $d_{1'}$) \\
        21 (e.g.,  $\nu^\text{isc}_{21}$ or $d_{21}$) & $2^\prime$ (e.g., $\nu^\text{isc}_{2'}$) & $3^\prime$ (e.g.,\, $d_{3'}$) \\
        \bottomrule
    \end{tabular}
\end{table}